\documentclass[a4paper,USenglish,cleveref,autoref,thm-restate]{lipics-v2021}

\hideLIPIcs  

\newcommand{\eps}{\varepsilon}

\newcommand{\gitrepo}[0]{\url{https://gitlab.com/rahmannlab/cuckoo-filters}}

\title{Smaller and More Flexible Cuckoo Filters}

\author{Johanna Elena Schmitz\footnote{
\label{contribution}Contributed equally.}}
{Algorithmic Bioinformatics, Faculty of Mathematics and Computer Science, Saarland University \and Saarbrücken Graduate School of Computer Science \and Center for Bioinformatics Saar, Saarland Informatics Campus, Saarbrücken, Germany}
{jschmitz@cs.uni-saarland.de}{https://orcid.org/0009-0002-6377-2561}
{}

\author{Jens Zentgraf\,$^1$}
{Algorithmic Bioinformatics, Faculty of Mathematics and Computer Science, Saarland University \and Saarbrücken Graduate School of Computer Science \and Center for Bioinformatics Saar, Saarland Informatics Campus, Saarbrücken, Germany}
{zentgraf@cs.uni-saarland.de}{https://orcid.org/0000-0001-9444-2755}
{}

\author{Sven Rahmann\footnote{Corresponding author: rahmann@cs.uni-saarland.de}}
{Algorithmic Bioinformatics, Faculty of Mathematics and Computer Science, Saarland University \and Center for Bioinformatics Saar, Saarland Informatics Campus, Saarbrücken, Germany}
{rahmann@cs.uni-saarland.de}
{0000-0002-8536-6065}{}

\authorrunning{J. E. Schmitz, J. Zentgraf and S. Rahmann}

\Copyright{{Johanna Elena} Schmitz, Jens Zentgraf and Sven Rahmann}  



\nolinenumbers 

\EventEditors{John Q. Open and Joan R. Access}
\EventNoEds{2}
\EventLongTitle{42nd Conference on Very Important Topics (CVIT 2016)}
\EventShortTitle{CVIT 2016}
\EventAcronym{CVIT}
\EventYear{2016}
\EventDate{December 24--27, 2016}
\EventLocation{Little Whinging, United Kingdom}
\EventLogo{}
\SeriesVolume{42}
\ArticleNo{23}

\begin{document}
\maketitle

\begin{abstract}
Cuckoo filters are space-efficient approximate set membership data structures with a controllable false positive rate (FPR) and zero false negatives. 
Conventional Cuckoo filters store multi-bit fingerprints of keys in a bucketed Cuckoo hash table using $k + 3$ bits per key and an additional space overhead factor of at least $1.05$ to achieve an FPR of $2^{-k}$. 
Additionally, the number of buckets is restricted to a power of 2, which may lead to much larger space overheads.

We present two improvements of Cuckoo filters. 
First, we remove the restriction that the number of buckets must be a power of 2. 
Second, we reduce the space overhead factor of Cuckoo filters to $1.06 \, (1+2/k)$, by using overlapping windows instead of disjoint buckets to maintain the load threshold of the hash table, while reducing the number of alternative slots where any fingerprint may be found.
A detailed evaluation demonstrates that the alternative memory layout based on overlapping windows decreases the size of Cuckoo filters both in theory and practice. 
A comparison with the state-of-the-art Prefix and Vector Quotient filters shows that the reduced space overhead makes windowed Cuckoo filters the smallest filters supporting online insertions, with similarly fast queries, but longer insertion times.

Code and workflows are available at \gitrepo.
\end{abstract}

\section{Introduction}
\label{sec:introduction}

Probabilistic filters are space-efficient data structures for approximate fast set membership queries.
They have many practical applications, such as for database systems \cite{metwally_duplicate_2005}, networks \cite{geravand_bloom_2013,mitzenmacher_network_2004}, storage systems \cite{luo_lsm-based_2020} and sequence analysis in computational biology \cite{holley_bloom_2016, lemane_kmtricks_2022, medina_bloom_2023}.

A probabilistic filter represents a set $K$ of keys from a key universe $\mathcal{U}$ and supports at least two operations:
(1) inserting a key, and (2) querying if a key is contained in the set.
Instead of storing an exact representation of the set~$K$, the keys are fuzzily represented in such a way that queries for keys that were inserted into the filter are always correctly answered (no false negative results).
Queries about keys that were not inserted are correctly answered most of the time, but sometimes incorrectly positively.
The probability of obtaining a false positive answer is called the false positive rate (FPR).
The FPR is controlled by the user and related to the space requirements of the filter.
To represent an arbitrary set of keys $K$ with cardinality $|K|=n$ with an FPR of $\eps = 2^{-k}$, the theoretically optimal space requirement is $nk$ bits. 
Practical implementations of filters need $Cnk$ bits with an overhead factor $C > 1$ that differs between filter types.
Different filter types try to achieve small overhead $C$ and fast insertion and query times, often with different trade-offs between space and time efficiency.
In addition, some filter variations support additional operations, such as deletion of keys (dynamic filters) \cite{fan_cuckoo_2014, geil_quotient_2018,vqf_pandey}, counting the number of times a key was inserted (counting filters) \cite{fan_summary_2000,pandey_general-purpose_2017}, or even storing arbitrary values with the keys (probabilistic key-value stores) \cite{wood_improved_2019}.
Other variants optimize the space requirements for a fixed set of keys, where the complete set of keys has to be known before filter construction starts (static filters) \cite{graf_xor_2020,graf_binary_2022,dillinger_fast_2022}.
Subsequently, we give a brief overview of some common filters.

Bloom filters \cite{bloom_spacetime_1970} were first introduced in 1970 and are still in common use. 
Bloom filters, when optimally configured for an FPR of $2^{-k}$, store a bit array with $m := (1/\ln 2) \, nk \approx 1.443 \, nk$ bits.
Disadvantages of Bloom filters are the relatively high overhead of 44.3\% and slow insertion and query times.
Blocked Bloom filters \cite{putze_cache_2007} achieve faster insertion and query times at the cost of even larger space, and the recent BlowChoc filters \cite{schmitz_blocked_2025} keep the advantages of Blocked Bloom filters while reducing the space requirements, sometimes even below those of standard Bloom filters.

However, alternative filter designs often perform better than Bloom filters. Cuckoo filters \cite{fan_cuckoo_2014}, Morton filters \cite{breslow_morton_2020}, (Vector) Quotient filters \cite{geil_quotient_2018,vqf_pandey}, and Prefix filters \cite{even_prefix_2022} store a $k$-bit fingerprint in a hash table instead of distributed bits in a bit array.
This design allows for storing additional values together with the key fingerprints simply by using additional bits, giving us not only a probabilistic set membership data structure but also a probabilistic key-value store without additional effort.
The FPR is controlled by the fingerprint size, where larger fingerprints lead to smaller FPRs.
Different filter types in this class differ in their hash collision resolution strategies.
For example, Cuckoo filters resolve collisions using Cuckoo hashing \cite{pagh_cuckoo_2004, fan_cuckoo_2014}, and Vector Quotient filters use ideas from Robin Hood hashing \cite{vqf_pandey}.

Depending on the collision resolution strategy, the filters achieve different load thresholds. 
For a hash table with $s$ slots, of which $n$ slots are occupied, the load (or fill rate) $r$ is defined as $r := n / s$.
The load threshold (or maximum load factor) is the largest load~$r$ for which the filter can be successfully constructed with high probability. 
To achieve low space overhead, we need to design a data structure that has a high load threshold while at the same time limiting the number of possible positions a key's fingerprint may be stored, as each alternative position would increase the FPR if not compensated for by additional bits per key.

If the whole key set is known in advance, static filters, such as XOR filters \cite{graf_xor_2020}, (Bumped) Ribbon filters \cite{dillinger_ribbon_2021,dillinger_fast_2022}, or Binary Fuse filters \cite{graf_binary_2022}, achieve small overhead factors but are not suitable for streaming applications where the key set~$K$ is usually not known in advance.

All of the above filters, even the non-static ones, require knowledge of the cardinality $n=|K|$ (or an accurate estimate of it) prior to filter construction. 
In contrast, extendable filters, such as Infini filters \cite{dayan_infinifilter_2023}, Aleph filters \cite{dayan_aleph_2024}, dynamic Cuckoo filters \cite{chen_dynamic_2017}, or consistent Cuckoo filters \cite{luo_consistent_2019}, increase their capacity if more keys arrive than originally planned for, at the cost of a higher memory usage and/or higher FPR. 
Adaptive Cuckoo filters \cite{mitzenmacher_adaptive_2020} can react to false positive queries by rearranging fingerprints in the Cuckoo filter. To recognize a false positive, adaptive Cuckoo filters require the whole key set to be additionally stored in a hash table and are thus only practical in applications where the Cuckoo filter only acts as a fast pre-filter.

\begin{table}[t]\centering
\caption{
Overview of non-extendable filter types supporting online insertions.
The overhead factor $C=C(k) > 1$ specifies the required space per key; one needs $Cnk$ bits to store $n$ keys with an FPR of $2^{-k}$.
If a filter only supports a number of slots that is a power of $2$, $C(k)$ can vary between the given formula (best case, see Figure~\ref{fig:comp_filters}) and twice that value (worst case; here given as worst-case $C(8)$ for a typical $k=8$ scenario).
The number of cache misses for insertion/lookup is a proxy for insertion/lookup time.
}\label{tab:filtertypes}
\begin{tabular}{llccccc}
\hline
& Overhead factor $C(k)$, &  Worst  & \multicolumn{2}{c}{Cache misses for} \\ 
Name [Ref.] & best case (max load)  & $C(8)$ & insert & lookup \\
\hline
Bloom \cite{bloom_spacetime_1970} & 1.443  & 1.443 & $k$ & $k$  \\
Standard Cuckoo \cite{fan_cuckoo_2014} & $1.050 \,(1+3/k)$ ${}^{(\dagger)}$  & 2.887 & 1 to $M$ ${}^{(*)}$ & 2\\
Improved Cuckoo [this] & $1.057 \,(1+2/k)$  & \textbf{1.321} & 1 to $M$ ${}^{(*)}$ & 2\\
Quotient \cite{bender_dont_2012} & $1.053\,(1+2.125/k)$ ${}^{(\dagger)}$  & 2.665 & some ${}^{(**)}$ & some ${}^{(**)}$ \\
Vector Quotient \cite{vqf_pandey} & $1.075\,(1+2.914/k)$ ${}^{(\dagger)}$  & 2.933 & 2 & 2\\
Prefix  \cite{even_prefix_2022} & $(1+\gamma)\;(1+2/k) + \gamma / k$ ${}^{(\ddagger)}$  & 1.359 & $\ge  1$ $^{(***)}$ & $\ge 1$ $^{(***)}$\\
\hline
\end{tabular}
\vspace*{1ex}\begin{flushleft}\footnotesize
${}^{(\dagger)}$: The number of buckets/slots in the filter must be a power of~2. \newline
${}^{(\ddagger)}$:~The Prefix filter achieves the same FPR for different parameters $\gamma$, here $\gamma = 1/\sqrt{2\pi \cdot 25}\;$ \cite{even_prefix_2022}. \newline
${}^{(*)}$:~For Cuckoo filters, $M$ is the maximum random walk length during insertion. Higher values of~$M$ achieve lower overhead, but require more time. \newline
${}^{(**)}$:~For the Quotient filter, cache misses depend on the load factor; higher loads (lower overhead) mean more cache misses and more time. \newline
${}^{(***)}$:~For the Prefix filter, cache misses depend on the load factor and whether the spare (second level; used if bins in first level are full) is accessed. If the spare is not accessed, insert and queries require 1 cache miss. The number of cache misses in the spare depends on the chosen spare filter (e.g., Cuckoo, Vector Quotient or Blocked Bloom filter).
\end{flushleft}
\end{table}

\begin{figure}[t]\centering
\includegraphics[width=1.0\linewidth]{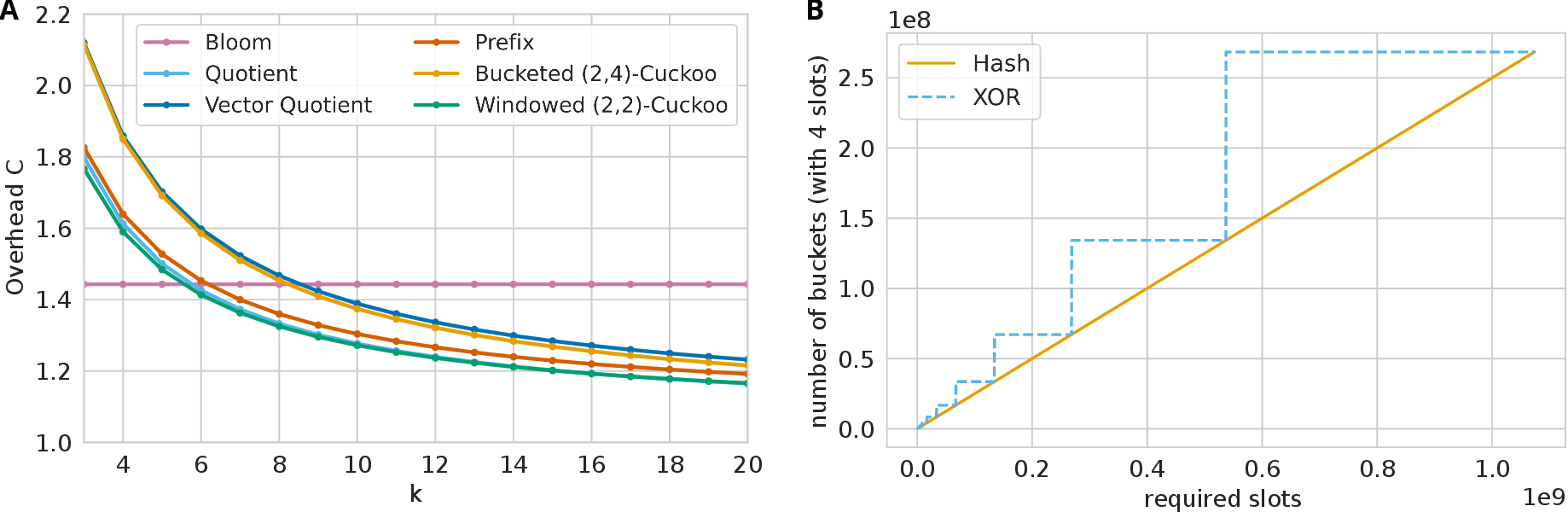}
\caption{
(A) Best-case overhead factors for different filters and different $k$ to get an FPR of $2^{-k}$.
The standard Cuckoo filter is the bucketed (2,~4) Cuckoo filter (orange).
(B) Number of required buckets in the original (2,~4) Cuckoo filter \cite{fan_cuckoo_2014} for a given desired number of slots (dashed blue staircase) vs.\ an optimally sized Cuckoo filter (orange diagonal):
In the original work, the number of buckets must be a power of~2, inducing large jumps in the size, resulting in additional space overhead.
}\label{fig:comp_filters}
\end{figure}

In this work, we focus on fingerprint-based non-extendable filters supporting online insertions; an overview is given in Table~\ref{tab:filtertypes}.
In contrast to the Bloom filter, the (best possible) overhead factor of the fingerprint-based filters decreases with increasing~$k$ (see also Figure~\ref{fig:comp_filters}), but many of these filters have additional technical or efficiency-related restrictions, which can increase the overhead by a factor of (almost)~2, and hinder flexible use.

We improve upon Cuckoo filters in two ways:
First, we remove a restriction that yielded low space overhead factors only when the number of inserted keys satisfied certain conditions.
Second, we change the memory layout to overlapping windows instead of non-overlapping buckets, obtaining a higher load threshold with fewer possible locations for each key, thus needing fewer bits of memory overall.
The advantages of the windowed memory layout also apply to other Cuckoo filter variants, such as the dynamic Cuckoo filter \cite{chen_dynamic_2017} or the adaptive Cuckoo filter \cite{mitzenmacher_adaptive_2020}, but we limit our evaluation to the standard Cuckoo filter with fixed capacity.
Our implementation is parallelized using independent subfilters, which can be filled by separate threads without locking or complex communication.
We thus obtain universally usable filters that are both small and fast.

After providing background on Cuckoo hashing and Cuckoo filters in Section~\ref{sec:background}, we describe how to improve Cuckoo filters in Section~\ref{sec:windowed_Cuckoo} and provide implementation details in Section~\ref{sec:implementation}. 
Section~\ref{sec:evaluation} contains a detailed evaluation of bucketed and windowed Cuckoo filters and compares them to other state-of-the-art filters, and Section~\ref{sec:conclusion} concludes.

\section{Background}
\label{sec:background}

\subsection{Multi-way Cuckoo hashing}
\label{sec:cuckoohashing}

In standard Cuckoo hashing \cite{pagh_cuckoo_2004}, a set of keys $K=\{x_1, x_2, \dots, x_{n}\}$ from a key universe $\mathcal{U}$ is stored (exactly) in a hash table with $s \ge n$ slots; indexed $0, 1, \dots, s-1$. 
After inserting $n$ keys, the load factor of the hash table is $r := n / s$.
The hash table may store additional data associated with each key.
Each key $x$ may be inserted into one of two possible slots, computed using two  hash functions $f_1: \mathcal{U} \to [s]:= \{0, \dots, s-1\}$ and $f_2: \mathcal{U} \to [s]$, randomly chosen from a universal family.
If both slots $f_1(x)$, $f_2(x)$ for key $x$ are already occupied, one of the keys at $f_1(x)$ or $f_2(x)$ is removed and re-inserted into its alternative slot. 
This slot might also be occupied, so a chain of removals and re-insertions starts until a free slot is found.
This can lead to long walks or end in a cycle, and the maximum load that can be achieved with a high probability in this setting is $1/2$, a rather low value.

\begin{table}[t]\centering
\caption{
Rounded theoretical load thresholds for $(2, l)$ Cuckoo hashing with buckets and windows, according to Walzer \cite{walzer_load_2023}.
}\label{tab:walzer_thresholds}
\begin{tabular}{lcccc}
\hline
layout  & $l=1$ & 2 & 3 & 4 \\
\hline
buckets (2-ary) & 0.5 & 0.8970118682 & 0.9591542686  & 0.9803697743\\
windows (2-ary) &  0.5 & 0.9649949234  & 0.9944227538 & 0.9989515932 \\
\hline
\end{tabular}
\end{table}

To achieve higher loads and consequently use less memory, Cuckoo hashing has been generalized in the following three ways.
\begin{enumerate}
\item In $d$-ary Cuckoo hashing, $d$ independent hash functions $f_1, f_2, \dots, f_d: \mathcal{U} \to [s]$ are used to compute $d$ candidate positions for each key \cite{fotakis_space_2005}.
\item In $(d, l)$~bucketed Cuckoo hashing, the hash table is divided into $B = \lceil s /l\rceil$ non-overlapping buckets, each containing $l$ slots. 
The $d$ hash functions $f_1,\dots,f_d: \mathcal{U} \to [B]$ pick alternative buckets.
Each key can be inserted into any of the $l$ slots inside the $d$ buckets \cite{dietzfelbinger_balanced_2007, zentgraf_fast_2021}.
\item $(d, l)$~windowed Cuckoo hashing works similarly to $(d, l)$~bucketed Cuckoo hashing but uses overlapping windows instead of non-overlapping buckets \cite{lehman_35-way_2009}. 
Each window with $l$ slots overlaps by $l-1$ positions with its surrounding windows.
The total number of windows is thus $W = s - l + 1$. 
Everything else works analogously to the bucketed version.
\end{enumerate}
Theoretical asymptotic load thresholds for these generalizations have recently been computed by Stefan Walzer \cite{walzer_load_2023}; see~Table~\ref{tab:walzer_thresholds} for relevant examples.
The load threshold increases with $d$ and $l$ and is higher for windows compared to buckets. 
However, larger $d$ leads to increased query times, since we search a key in each of the $d$ buckets or windows at different memory locations, each likely causing a cache miss, while a search inside a bucket or window is local.
Hence, we consider only $d=2$.
Since existing Cuckoo filter implementations use buckets, it is natural to ask whether a windowed memory layout reduces the memory overhead while retaining the fast query times of Cuckoo filters also in practice.

\subsection{Cuckoo filters}
\label{sec:cuckoo}

A Cuckoo filter is a probabilistic data structure that is based on Cuckoo hashing, introduced using $(2,4)$~bucketed Cuckoo hashing \cite{fan_cuckoo_2014}.
A Cuckoo filter uses a \emph{fingerprint} hash function $f_0: \mathcal{U} \to [2^{q}]$ that computes a $q$-bit fingerprint for a key.
Each key is assigned to two distinct \emph{buckets} $b_1$ and $b_2$ from a total of $B$ buckets.
A bucket consists of a constant number of available slots (4~in \cite{fan_cuckoo_2014}), and each slot may hold a fingerprint.
Therefore, the fingerprint of a key can be stored in any of $2\cdot 4 = 8$ slots.
To guarantee an FPR of $2^{-k}$, with 8 possible locations for each key, the fingerprint size needs to increase to $q := k+3$ bits, yielding an FPR bounded by $8 \cdot 2^{-(k+3)} = 2^{-k}$.

Since only the fingerprint is stored, the two bucket addresses must be restricted so that one can be calculated from the other and the fingerprint. 
If we do not want to sacrifice a bit that remembers whether the fingerprint is stored in its first or second bucket, we must use a symmetric function to compute both buckets.
The standard Cuckoo filter solves this problem as follows \cite{fan_cuckoo_2014}.
A hash function $f_1: \mathcal{U} \to [B]$ computes the first bucket address of a key~$x$, given by $b_1 = f_1(x)$. 
The second bucket address $b_2$ is given by the bit-wise XOR of a hash of the fingerprint $h(f_0(x))$ and the first bucket address $b_1$, where $h:[2^q] \to [B]$.
Conversely, the first bucket address is obtained back by XORing $b_2$ with the same fingerprint hash.
The XOR operation is only guaranteed to return a valid address if the number of buckets is a power of~2.
A proposed simplification \cite{eppstein_cuckoo_2016} uses the $q$-bit fingerprint $f_0(x)$ directly instead of a hash value $h(f_0(x))$.

\subparagraph{Insertions}
Inserting a key~$x$ with 2~hash functions and a bucket size of~$l$ works as follows.
If any of the $2l$ slots in which the fingerprint of~$x$ can be stored is empty, it is stored there, giving priority to slots in the first bucket $b_1$.
If all $2l$ slots are full, pick a random one of these slots, remove the stored fingerprint, say, of key $x'$, and insert fingerprint $f_0(x)$ at the now free slot.
Then, attempt to insert the removed fingerprint $f_0(x')$ into its alternative bucket (which can be computed from its former address and the fingerprint itself, without knowledge of $x'$).
This procedure is repeated for up to a given number of steps.
The insertion fails if no free slot can be found. In this case, the filter size must be increased and all insertions repeated.
The probability of failure approaches zero if the number of buckets is large enough (and enough steps are allowed).
In the described $(2, 4)$ configuration, $B=1.02 \cdot n/4$ buckets, or $1.02\, n$ slots are sufficient in theory to insert all $n$ elements.
In practice, to limit the length of the random insertion walks, the number of buckets and slots is chosen larger, and the original work \cite{fan_cuckoo_2014} uses $1.05\, n$ slots.
It has to be noted that a Cuckoo filter (if not over-provisioned) cannot tolerate many additional insertions; these will simply fail.

\subparagraph{Queries} 
To query the presence of a key in a $(2,4)$~bucketed Cuckoo filter, search all eight possible slots and return \texttt{True} if and only if the fingerprint $f_0(x)$ was found at any of the slots. 
Queries are fast because at most two distinct memory locations (buckets $b_1$ and $b_2$) need to be accessed; the remaining memory accesses are local within each bucket.

\section{Flexible windowed Cuckoo filters}
\label{sec:windowed_Cuckoo}

We improve upon the standard Cuckoo filter in two ways:
First, we introduce a different way to find the alternative address from the given one and the fingerprint, using a signed offset.
Second, we use overlapping windows instead of disjoint buckets.

\subsection{Moving between alternative buckets or windows}

In the standard Cuckoo filter \cite{fan_cuckoo_2014}, there are two alternative buckets to store the $q$-bit fingerprint $f_0(x)$ of key~$x$.
If $b$ is one of the buckets, $b'$ is the alternative bucket, and $h: [2^q] \to [B]$ is a hash function, then setting $b' = b \oplus h(f_0(x))$ is symmetric in $b, b'$ but requires that the number of buckets satisfies $B=2^\beta$ for some integer $\beta$ to guarantee valid bucket addresses $b, b' \in [B]$.
This may introduce a significant memory overhead (up to an additional factor of 2.0; see Figure~\ref{fig:comp_filters}B).
An alternative \cite{graf_xor_2020} proposed $b' := (B - (b + f_0(x))) \bmod B$, which is also symmetric in $b, b'$ and imposes no restrictions on~$B$.
However, it does not allow further randomization, which makes it vulnerable to adversarial input data and introduces a dependency between the alternative locations $b$ and $b'$.

We propose a different approach, consisting of the first bucket hash function $f_1: \mathcal{U} \to [B]$ and an offset hash function $f_2: [2^q] \to [B-1]$, mapping any fingerprint to an offset less than~$B$.
For a key $x$, we have the first bucket address $b = f_1(x)$ and $b' := (b + f_2(f_0(x))) \bmod B$ with $b' \ne b$. 
From $b'$, bucket $b$ cannot be reconstructed symmetrically, but asymmetrically as $b = (b' - f_2(f_0(x))) \bmod B$.
Hence, we need to use a bit in addition to the fingerprint to store the current bucket choice ($b$ or $b'$) that tells us whether we need to add or subtract the offset to obtain the alternative bucket.

It seems disadvantageous that we need one extra bit in each slot.
However, when we query the presence of a key, we only consider the key to be present if both the fingerprint and the choice bit are identical.
Therefore, we can use the choice bit as one of the extra bits needed to counteract the multiple possible slots. 
We do \emph{not} require that this bit is uniformly distributed between 0 and 1 across the filter; this can be seen by the following argument.

Let $T=2^t$ be the number of possible slots for any fingerprint.
Assume that fraction~$p$ of all fingerprints (of length $q=k+t$) are stored in their respective first bucket with choice bit~0, and the remaining fraction $1-p$ in their second bucket with choice bit~1.
A false positive occurs if we find the fingerprint together with choice bit~0 in the first bucket (probability $\le T \cdot 2^{-q} \cdot p$) or with choice bit~1 in the second bucket (probability $\le T \cdot 2^{-q} \cdot (1-p)$).
The total probability is thus bounded by $T \cdot 2^{-q} \cdot (p + 1-p) = 2^{t-q} \cdot 1 = 2^{-k}$, independently of~$p$.

\subsection{Windowed layout}

\begin{figure}\centering
\includegraphics[width=0.85\linewidth]{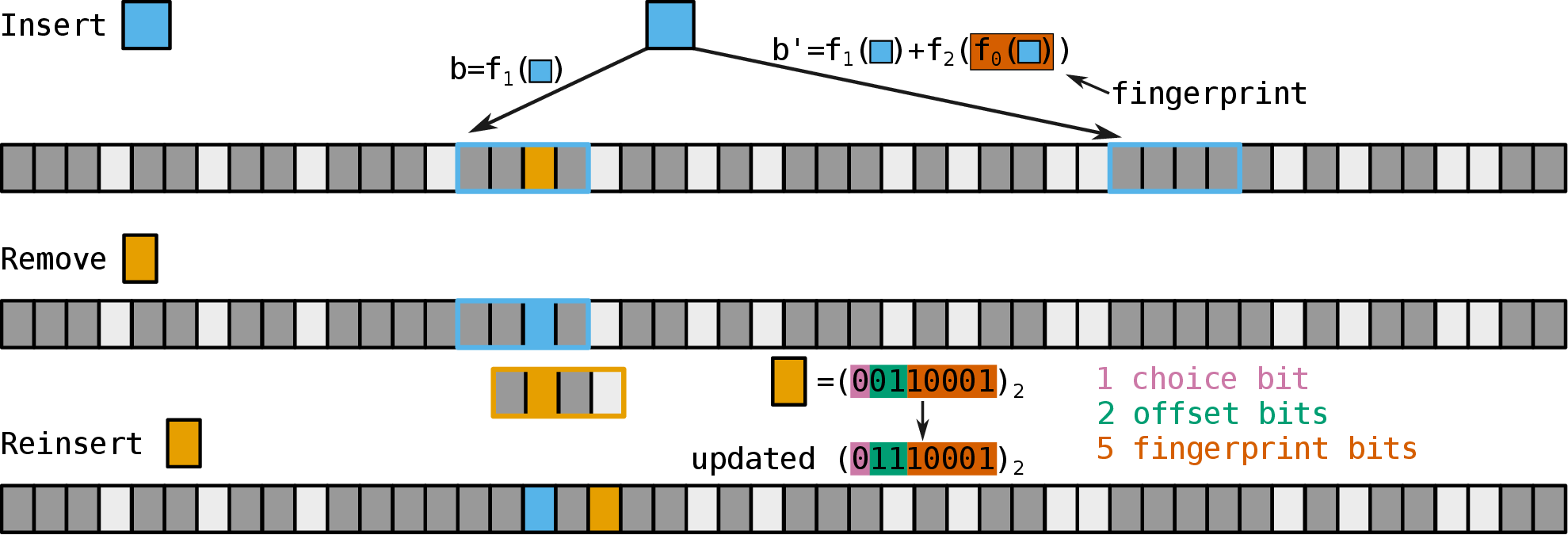}
\caption{
Inserting a key's fingerprint into a Cuckoo filter with window size~4 and an FPR of $2^{-5}$:
Compute both window addresses and search for an empty slot in either window. 
If both windows are full, as shown here, remove a random fingerprint from one of the windows (here, the orange fingerprint), compute its two possible windows based on the choice and offset bits, and try to re-insert the fingerprint at one of the alternative slots.
Here, the removed fingerprint can directly be re-inserted into its current window, since the last position in the window is still empty.
}\label{fig:example_windowed_cuckoo}
\end{figure}

The standard Cuckoo filter uses $(2,4)$ bucketed Cuckoo hashing \cite{fan_cuckoo_2014}.
Alternatively, the array may be divided into $W$ overlapping windows of size $l$, where each window overlaps with the previous window by $l-1$ slots.
Using windows has been shown to increase the load threshold for Cuckoo hash tables in comparison to buckets \cite{walzer_load_2023,lehman_35-way_2009}; see Table~\ref{tab:walzer_thresholds}.
This allows us to move from $(2, 4)$ bucketed Cuckoo hashing to $(2, 2)$ windowed Cuckoo hashing while maintaining a comparably high load threshold in practice; theoretically, it drops from $0.980$ to $0.965$.

However, new complications arise: Any slot of the hash table now belongs to $l$~different windows, and from the location alone we do not know the window number.
Hence, we also need to store the offset from the window start, a number in $[l]$, taking $\log_2 l$ bits.
(For convenience, $l$ should therefore be a power of~2, such as 2 or 4.)
Fortunately, again, these need not be extra bits but can be used as part of the bits that counteract the FPR multiplier.
For concreteness, in a $(2, 2)$ windowed Cuckoo filter, we use a $k$-bit fingerprint, one additional choice bit to indicate which of the two alternative window locations we are using, and one additional offset bit within the window (first or second slot).
Similarly to the choice bit, the distribution of the window offset bits need not be uniform over all $2^l$ possible values.

Inserting or querying a key works analogously to the version with buckets. 
To insert a key, check if any of the $2l$ slots is empty. 
If yes, insert the fingerprint into the empty slot.
Otherwise, remove a random fingerprint and try to re-insert it into an alternative slot. 
To re-insert the removed fingerprint, compute both its current and alternative window based on the stored bits (Figure \ref{fig:example_windowed_cuckoo}).
Run this loop of removal and re-insertion until an empty slot is found, but at most for a fixed number of steps.
For queries, return \texttt{True} if and only if any of the $2l$ possible slots contain the correct fingerprint and the correct choice and window offset bits for the currently searched slot.

In summary, we need to store $k+2$ bits per key in a $(2, 2)$ windowed Cuckoo filter for an FPR of $2^{-k}$, and in practice use a maximum load of $\approx 0.94$ (see Figure~\ref{fig:overhead}). 
The space overhead of a $(2, 2)$ windowed Cuckoo filter is therefore $1/0.94 (1+2/k) \approx 1.06 (1+2/k)$.
The main differences with regard to the standard Cuckoo filter \cite{fan_cuckoo_2014} are that in our work,
\begin{enumerate}
\item the additional bits are not (randomized) extensions of the fingerprint but have meaning as choice bit and offset bits,
\item we reduce the number of offset bits from 2 to 1, saving space, because the load threshold for Cuckoo hashing with two slots in windows and for four slots in buckets are comparable ($0.965$ vs.\ $0.980$).
Filling the hash tables slightly below their theoretical load thresholds, the overhead factor is reduced from $1.05 (1+3/k)$ in \cite{fan_cuckoo_2014} to $1.06 (1+2/k)$, e.g.,\ from $1.365$ to $1.272$ for $k=10$ (an FPR of $1/1024$).
\end{enumerate}

\section{Implementation Details}
\label{sec:implementation}

We now discuss our implementation using optimized just-in-time compiled Python, which hash function we use, how we distinguish empty from full slots in the hash table, how we execute bit-parallel queries over several slots, how we perform prefetching, and how we parallelize insertion.

\subsection{Just-in-time compilation}
\label{sec:jit}

We have implemented the improved Cuckoo filters for integer keys in just-in-time compiled Python using the \texttt{numba} package \cite{numba} with typed \texttt{numpy} arrays. 
This offers several benefits, such as highly optimized machine code, the use of LLVM intrinsics, and the option to choose parameters during runtime but before compilation. 
For example, this allows us to provide all parameters of the hash functions as compile-time constants for additional optimizations, achieving speeds comparable to compiled \texttt{C} or \texttt{C++} code, or sometimes even faster, due to the increased possibilities for optimization.
Since most state-of-the-art filters are implemented in compiled languages, we compared our reference implementation of the original Cuckoo filter with an existing \texttt{C++} implementation \cite{fan_cuckoo_2014}; see Appendix~\ref{app:numba}.
Since both implementations achieve similar running times, using just-in-time compiled Python comes with little penalties and allows us to compare our implementation with other state-of-the-art filters written in low-level languages.

\subsection{Hash functions}
\label{sec:hash_functions}

The hash functions for addressing the subfilter, the buckets or windows, and for computing the fingerprints do not need to satisfy particularly strong properties, but they need to be reasonably fast to compute and have sufficiently many free parameters that can be chosen.
In our implementation, we expect $u$-bit unsigned integer keys~$x$ with $u\le 64$ and use linear hash functions of the form $x \mapsto (ax) \bmod b$, where $a$ is a randomized integer, and $a$ and $b$ are coprime.
Note that there is always an implicit mod $2^{64}$ operation after every operation due to the unsigned 64-bit arithmetic; so in fact, we are computing $x \mapsto ((ax) \bmod 2^{64}) \bmod b$.
Modifications for a more uniform distribution of the hash values are applied if $b$~is divisible by~2 or even a power of~2:
If~$b$ is a power of~2, we only consider the $\log_2 b$ most significant bits of $ax$ and use $x\mapsto ((ax) \gg (u - \log_2 b)) \bmod b$.
If $b$ is even but not a power of~2, we perform a bitwise XOR operation between the $q := \lfloor u/2 \rfloor$ most significant bits and the $u/2$ least significant bits of $(ax)$ first, resulting in $x \mapsto ((ax) \oplus ((ax) \gg q)) \bmod b$.

\subsection{Distinguishing full slots from empty slots}
\label{sec:emptyslots}

We have so far not discussed how we can distinguish an empty slot from a full slot.
The hash function $f_0: \mathcal{U} \to [2^k]$ maps a key~$x$ to its $k$-bit fingerprint $f_0(x)$. 
We use one of the possible fingerprint values (by convention $0$) to indicate an empty slot.
Hence, we reduce the set of valid fingerprints and instead use a hash function $f'_0: \mathcal{U} \to [2^{k} - 1]$. 
The fingerprint is given by $f_0(x) := f'_0(x) + 1$. 
The loss of one fingerprint value leads to a higher FPR of $1/(2^k-1)$, the probability that two random fingerprints collide. 
Although this effect is measurable for small~$k$, it can be considered negligible for typically used larger values of $k$; see also Figure~\ref{fig:overhead} in Section~\ref{sec:eval_fpr}.
The same compromise was made for the original Cuckoo filter \cite{fan_cuckoo_2014}.

\subsection{Bit-parallel engineering}
\label{sec:engineering}

We adapt the bit-parallel optimizations from Fan et al. \cite{fan_cuckoo_2014} for windowed Cuckoo filters for sufficiently small~$k$, where several slots fit into a single 64-bit integer.
We discuss them by example using $(2,4)$ windowed Cuckoo filters with $k=5$, which has $q = k+3 = 8$-bit slots.
For other configurations, similar ideas are implemented with appropriate modifications.
For large~$k$, we examine each slot separately at the cost of throughput.

We load the contents of a window into a 64-bit integer register that we (conceptually) partition into four adjacent slots (for 32 bits, plus 32 unused bits).
To check whether there is an empty slot (all zeros), we use two pre-computed bit masks, \texttt{lo} and \verb|hi = lo << (q-1)|, indicating the least significant and most significant bits in each slot, respectively.
By computing the bit pattern \verb|e := (x - lo) & (~x) & hi|, the 1-bits of \texttt{e} are those 1-bits of \texttt{hi} that indicate which slots are empty.
If \texttt{e} is non-zero, there exists an empty slot, and its index (from the right) can be computed by dividing the number of trailing zeros (obtained by a \texttt{cttz} LLVM intrinsic) of~\texttt{e} by~\texttt{q}.
In the example, the result is nonzero and has 7 trailing zeros, indicating that slot 0 in $x$ is empty.
Comparing all 4 slots at the same time in this bit-parallel manner saves the overhead of a loop or separate comparisons.

\begin{table*}[!h] 
{\footnotesize\protect\begin{verbatim}
                slot        3 |        2 |        1 |        0
                hi = 10000000 | 10000000 | 10000000 | 10000000   high bits in each slot
                 x = 10001111 | 00000000 | 11100110 | 00000000   window contents x
                lo = 00000001 | 00000001 | 00000001 | 00000001   low bits in each slot
            x - lo = 10001101 | 11111111 | 11100100 | 11111111   -: subtraction
                ~x = 01110000 | 11111111 | 00011001 | 11111111   ~: bit-wise inversion
       (x-lo) & ~x = 00000000 | 11111111 | 00000000 | 11111111   &: bit-wise and
  (x-lo) & ~x & hi = 00000000 | 10000000 | 00000000 | 10000000   result e, 7 trailing 0s
\end{verbatim}}
\vspace{-4ex}\end{table*}

Similarly, to search for a fingerprint in any of the four slots, we can create a single bit mask that contains the valid bit patterns for each slot.
For $(2, 4)$ windowed Cuckoo filters, we use one choice bit and two window offset bits.
Assume that the 5-bit fingerprint is \verb|11100|, that we are examining window choice 1, and that the offset bits increase with decreasing slot number.
Then the valid slot bit patterns can be encoded in a mask \texttt{y}, and we can check if the bit patterns agree in one of the slots (and which one) by taking the bit-wise XOR \verb|x^y| and checking for a zero slot like above. The two possible masks for \texttt{y} that encode the valid bit pattern for the first and second window choice are precomputed based on $d$ and $l$ and used as compile-time constants in the lookup function.
In the following example, we find the fingerprint in slot~1.
\begin{table*}[!h] 
{\footnotesize\begin{verbatim}
      slot        3 |        2 |        1 |        0
  bit type CWWFFFFF | CWWFFFFF | CWWFFFFF | CWWFFFFF   choice/window offset/fingerprint
       y = 10011100 | 10111100 | 11011100 | 11111100   valid bit pattern per slot
       x = 10001111 | 00000000 | 11011100 | 00000000   window contents x
   x ^ y = 00010011 | 10111100 | 00000000 | 11111100   ^: XOR
\end{verbatim}}
\vspace{-4ex}\end{table*}

\subsection{Software prefetching}
\label{sec:prefetching}

During Cuckoo filter operations, a considerable amount of processing time is spent waiting for the RAM to fetch the data. 
Since we have random memory access patterns when querying a Cuckoo filter, automatic hardware prefetching, which is based on regular access patterns, is not helpful.
Instead, for batch queries, we may use explicit software prefetching to reduce the number of cache misses.
When queries come in batches, we may look ahead a number of queries and already prefetch the addresses of the slots that we will need in the future from RAM into the level-1 cache, using the LLVM \texttt{prefetch} intrinsic. 
The lookahead offset can be provided by the user as a compile-time constant; the optimal offset depends on the hardware (see Section~\ref{sec:benchmark_prefecth} and Appendix~\ref{app:prefetch}).

\subsection{Parallelization}
\label{sec:parallelization}

We parallelize the filter as follows: 
We create $F$ independent subfilters such that the total number of buckets (or windows) $B$ is divided equally among the $F$ subfilters, with $\lceil B / F \rceil$ buckets or windows per subfilter. 
Each subfilter is a cache-line-aligned bit array.
For inserting or querying a key, the subfilter index is computed by an additional hash function $f_\text{sub}: \mathcal{U} \mapsto [F]$. 
The bucket (or window) address hash function now maps the key $x$ to any of the $\lceil B / F \rceil$ buckets (or windows) in the subfilter. 
To insert keys into the subfilter, one main thread delegates the work to one thread per subfilter, responsible for performing the subfilter insertions.
To avoid locks, the threads communicate via message buffers using a consumer-producer principle, as in \cite{hackgap}.
The main thread computes the subfilter number $f_\text{sub}(x)$ for each key $x$ and copies the key to a buffer assigned to this subfilter.
For each subfilter, several such buffers exist, and these buffers are only read by one consumer thread responsible for insertions into this particular subfilter.
If a buffer is full, the main thread sets an atomic volatile \texttt{ready\_to\_read} flag and continues inserting keys into other buffers.
Among the buffers assigned to a specific subfilter, the corresponding consumer thread searches for a buffer marked as \texttt{ready\_to\_read} and inserts the keys into the subfilter.
After all keys in the buffer are inserted, the consumer thread marks the buffer as \texttt{ready\_to\_write} indicating to the main thread that it can again insert keys into the buffer.

Queries that are processed after all keys have been inserted only need read access.
No synchronization between query threads is needed, and queries are distributed in batches evenly across threads.

Assigning the keys randomly to $F$~independent subfilters causes a randomly varying load factor per subfilter.
However, this effect is negligible for large enough~$n$ because the number of keys in one subfilter follows a Binomial distribution with mean $\mu=np$ (where $p=1/F$) and standard deviation $\sigma=\sqrt{np(1-p)}$. 
Since we already use a space overhead of 6\% ($0.06\, \mu$), the additional overhead used by, say, $4\sigma$, is negligible for large $n$.
As an example, consider $n=10^8$ keys with $F=10$ subfilters, so $p=1/10$, and the mean number of keys per subfilter is $\mu = 10^7$ with $\sigma = 3000$.
Space overhead per subfilter (6\%) is $600\,000$ slots, while $4\sigma =12\,000$ slots, which disappear among the 6\%.
Hence, for large~$n$, parallelization does not incur a significant additional space overhead compared to the single-threaded version.

\section{Evaluation}
\label{sec:evaluation}

After describing the experimental setup (Section~\ref{sec:setup}), we start by comparing bucketed with windowed Cuckoo filters.
We first compare the theoretical load threshold against the practically achievable load for different filter configurations (Section~\ref{sec:eval_fill_rate}).
Next, we evaluate the time-memory trade-off for increasing loads (Section~\ref{sec:eval_time_memory}). 
In Section~\ref{sec:eval_fpr}, we evaluate the actual FPRs, loads, and overhead factors for different desired FPRs at target loads of $0.98 \cdot T$, where $T$ is the theoretical load threshold for a given configuration (Table~\ref{tab:walzer_thresholds}).
Performance benchmarks of our implementation are shown in Section~\ref{sec:eval_time} and performance using software prefetching is shown in Section~\ref{sec:benchmark_prefecth}.
Then, we compare the throughput of our Cuckoo filter implementations with two existing implementations of state-of-the-art filters, the Vector Quotient filter and the Prefix filter (Section~\ref{sec:eval_throughput}).
We omit static filters in order to focus on filters used in applications where the whole key set is not known in advance but where we have a good estimate of its size, which is typical in genome research applications in bioinformatics.
In addition, we exclude other filters, like Bloom filters, Morton filters, or Quotient filters, that showed worse results compared to Prefix and Vector Quotient filters in previous analyses \cite{even_prefix_2022, vqf_pandey}.

\subsection{Experimental setup.}
\label{sec:setup}

We evaluate the original Vector Quotient filter implementation from \cite{vqf_pandey} (downloaded from \url{https://github.com/splatlab/vqf}, written in \texttt{C++}) and the original Prefix filter implementation \cite{even_prefix_2022} (from \url{https://github.com/TomerEven/Prefix-Filter}, written in \texttt{C++}) using a Cuckoo filter or VQF as a spare.

Evaluations were run on a PC workstation with an AMD Ryzen 9 9950X CPU with 16 cores and hyperthreading and 64 GB of DDR5 memory (6000 MHz, CL40). 
All benchmarks were performed on random unsigned 64-bit integer keys generated with \texttt{numpy}.
Reported times are wall times, including just-in-time compilation and excluding data load time, except for multithreaded insertion, where reader and inserter threads load and insert keys in parallel.
We report averages over five runs.

For insertions, we perform lookup-and-insert operations, i.e., we only insert a key if it is not already classified as present in the filter, except when comparing the throughput between filters in Section \ref{sec:eval_throughput}.

\subsection{Comparison of load thresholds with achievable loads}
\label{sec:eval_fill_rate}

\begin{figure}[t]\centering
\includegraphics[width=1.\linewidth]{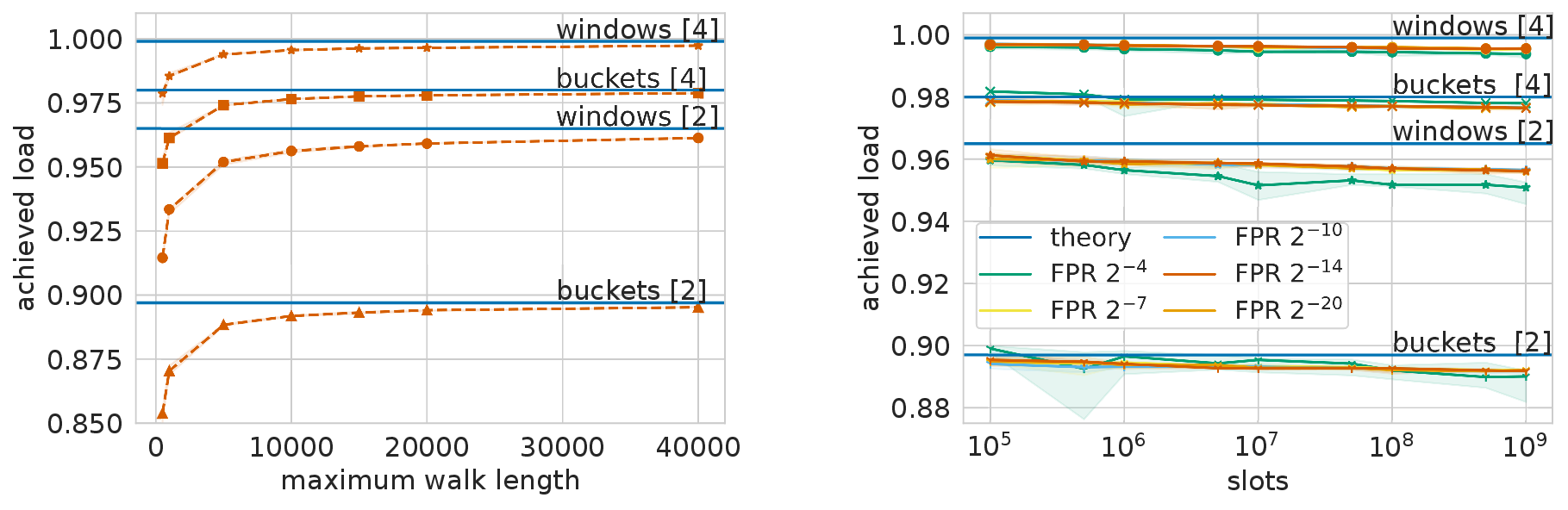}
\caption{
Comparison of theoretical load thresholds (horizontal blue lines), using windows and buckets and $l\in\{2,4\}$ slots per bucket or window, with achievable loads. 
Left:
Achievable loads for different maximum random walk lengths (x-axis) for the different types of Cuckoo filters with an FPR of $2^{-14}$ and $10^9$ slots.
Right:
Achievable loads for different FPRs (color) and filter sizes (x-axis) using a fixed random walk length of 10\,000.
}\label{fig:max_fill}
\end{figure}

We created datasets of random 64-bit integers and counted the number of inserted keys until the insertion failed for the first time. 
We tabulated the percentage of filled and empty slots in the filter to compute the achievable load of bucketed and windowed Cuckoo filters with different parameters.
We show the average (main line), the minimum and maximum (shaded region) over five runs.

\subparagraph{Random walk length}
Figure~\ref{fig:max_fill}~(left) shows the achievable loads for different maximum random walk lengths.
As expected, the achievable load increases for longer random walk lengths and converges to the theoretical load threshold if enough steps are allowed for the random walk during insertion.
We conclude that 10\,000 steps are enough to achieve loads close to the thresholds with high probability. 
Note that large numbers of steps are only taken when the filter is almost at capacity; as long as it is sufficiently empty, insertion happens in a single step (or with a few steps), independently of the maximum walk length (see Appendix~\ref{app:random_walk_distribution}).

\subparagraph{Achievable loads for different FPRs and filter sizes} 
For a fixed random walk length, the achievable load should be independent of the FPR and the relative filter size.
In practice, the achievable load appears independent of the FPR, except for small~$k$ with, e.g., an FPR of $2^{-4}$, for which the achieved load is lower than for higher~$k$ (Figure~\ref{fig:max_fill} right).
A possible explanation is that having only $2^4-1 = 15$ distinct fingerprints reduces the randomness for the second bucket or window. 
In addition, the achievable load decreases slightly for larger filters (with constant overhead factor, i.e., both $n$ and the filter size are at constant ratio), which may also be explained by the problem of mapping the small space of fingerprints to the much larger number of buckets or windows. 

\subsection{Time-memory trade-off}
\label{sec:eval_time_memory}

\begin{figure*}[t]\centering
\includegraphics[width=1.0\textwidth]{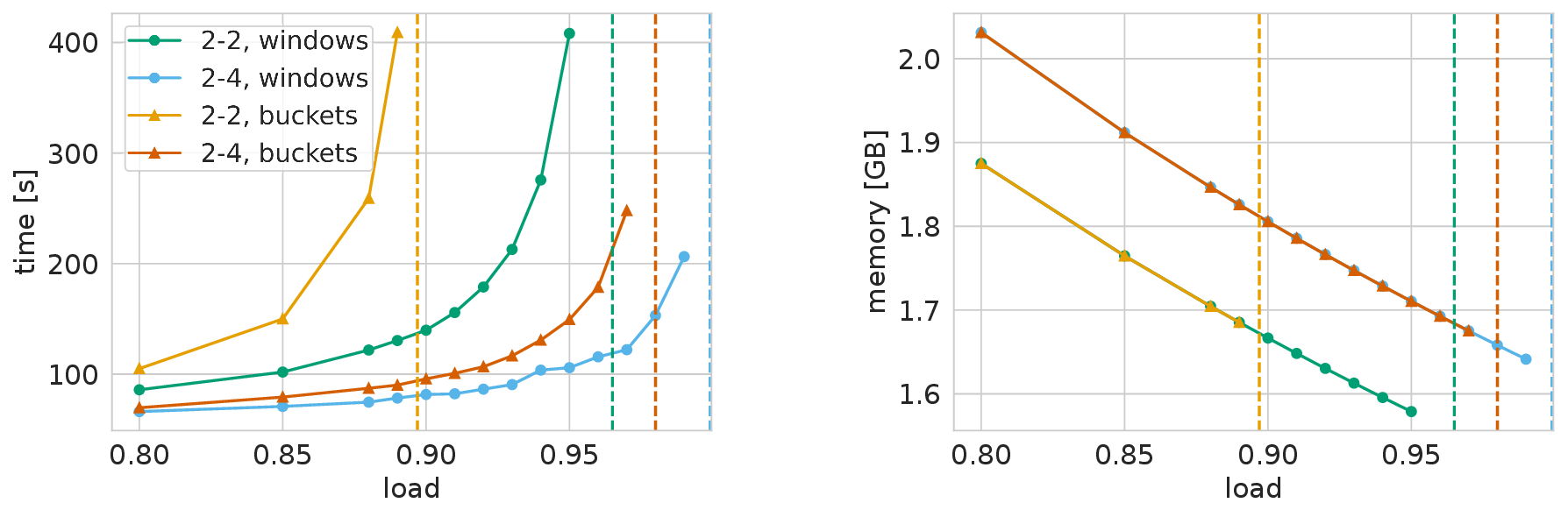}
\caption{
Time (left) and memory (right) requirements for inserting a billion ($10^9$) keys into different Cuckoo filter types (distinguished by color) at different loads (x-axis) for a fixed FPR of $2^{-10}$ and a fixed maximum random walk length of 10\,000. 
Dashed vertical lines correspond to load thresholds.
Memory use for windows and buckets coincides, i.e., below their respective thresholds, the yellow and green lines overlap, as well as the blue and orange lines.
}\label{fig:time_vs_fill}
\end{figure*}

Figure \ref{fig:time_vs_fill} shows the time and memory requirements for inserting a billion keys with a fixed FPR of $2^{-10}$ at different final load factors.
Higher loads result in higher insertion times, with steep increases as the load threshold is approached.
For low loads of 0.8, insertion times are comparable.
At load 0.9, already $(2,2)$ buckets are impossible, and the time for $(2,2)$ windows has increased from 85 seconds to 140 seconds, but the time for both $(2,4)$ configurations has only slightly increased from 65 seconds to 94 seconds at most.
Windowed Cuckoo filters have higher load thresholds compared to bucketed Cuckoo filters with the same number of slots per bucket/window.
Therefore, at the same load, the windowed $(d,l)$ Cuckoo filter is faster than the  bucketed version.

The memory requirements depend on the window size or bucket size and the load, but there is no difference between windowed and bucketed versions.
At the same load, Cuckoo filters with a window size or bucket size of~2 are smaller compared to Cuckoo filters with a size of 4 because one less bit per slot is needed. 
Windowed Cuckoo filters achieve higher loads and faster insertion times than their bucketed counterparts at fixed load, so they are an overall improvement.

\subsection{Actual overhead factors, loads, and FPRs}
\label{sec:eval_fpr}

\begin{figure*}[t]\centering
\includegraphics[width=1.\textwidth]{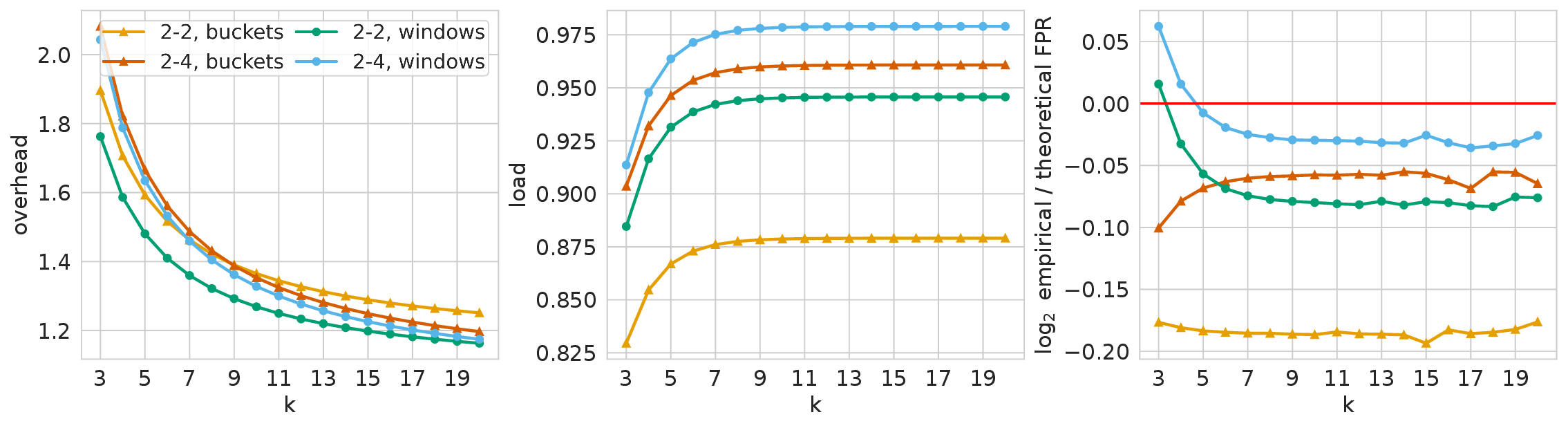}
\vspace*{-2ex}
\caption{
Comparison of actual properties (memory overhead, actual load, empirical FPR) of Cuckoo filter types (distinguished by color) designed for FPRs of $2^{-k}$ for varying $k$ (x-axis), using a fixed maximum random walk length of 10\,000, with $n=2\times 10^9$ keys and 5~subfilters for parallelization.
Insertions are performed only if the filter does not already report a key as present.
Left: Actual memory overhead factor $C$ for a Cuckoo filter with $Cnk$ bits for a target FPR of $2^{-k}$; lower is better. 
Middle: Empirical loads.
Right: Empirical FPRs, relative to the target FPR of $2^{-k}$, as log ratios. Points below the horizontal red line at 0.00 have an actual FPR that is better than the target FPR.
}\label{fig:overhead}
\end{figure*}

According to the observations of Section~\ref{sec:eval_fill_rate}, we choose a filter size that ensures a load of at most 98\% of the theoretical load threshold and a maximum walk length of 10\,000 steps. 
However, the actual random walk length is usually much lower; see Appendix~\ref{app:random_walk_distribution}.
Using $n=2\cdot 10^9$ random 64-bit integer keys, we now compare the resulting actual memory overhead factors, empirical FPRs, and observed load factors for different values of~$k$ (Figure~\ref{fig:overhead}).
The actual memory overhead factor is obtained by dividing the actual number of used bits by $nk$.
The actual load is the number of occupied slots divided by the number of total slots.
The empirical FPR is obtained by querying $n' = 2\cdot 10^9$ random keys that were not previously inserted into the filter, computed as the number of times the lookup erroneously returned \texttt{True}, divided by $n'$. 

\subparagraph{Overhead factors} 
For all Cuckoo filter variants, the overhead factor decreases for increasing~$k$ (i.e., for lower FPRs; see Table~\ref{tab:filtertypes} and Figure \ref{fig:overhead} left).
Windowed Cuckoo filters with window size 2 have the lowest overhead factors, since a window size of 2 saves one bit per slot compared to buckets or windows of size 4, and they achieve high load thresholds.
Although buckets of size 2 use the same number of bits per slot, their load threshold is lower, and they therefore have higher overhead factors compared to windows of size~2.
For small~$k$, buckets of size~2 have a lower overhead factor compared to windows of size 4, but from $k \ge 8$, the overhead factor for windows of size 4 is lower, since the higher load compensates for the additional bit per slot.
Cuckoo filters with buckets of size $4$ have higher overhead factors compared to windows of size $4$ due to the higher load threshold.

\subparagraph{Empirical loads}
As shown in Figure~\ref{fig:overhead} (middle), for $k \ge 8$, the empirical load is independent of~$k$ and corresponds to the targeted load of 98\% of the load threshold. 
For smaller $k$, the observed load is actually lower. 
Since small fingerprints have high collision probabilities (of $1/(2^k - 1)$), several keys are reported as already present during the performed lookup-and-insert operations and not inserted again, causing an overall lower load.

\subparagraph{Empirical false positive rates}
We target an FPR of $2^{-k}$ by using $k+2$ or $k+3$ bits per slot, for a window and bucket size of 2 or 4, respectively.
In practice, several other effects increase or decrease the actual FPR.
Figure~\ref{fig:overhead} (right) shows log-ratios between empirical FPR and $2^{-k}$.
For $k\ge 5$, the observed FPR is lower than $2^{-k}$ because the filter is not full.
Actual non-present keys have a small chance of hitting an empty slot instead of a stored fingerprint, decreasing the FPR.
Since the load decreases for small~$k$ (see above), the observed FPR should also decrease for small $k$.
However, for window-based filters, the FPR is higher for $k \leq 4$:
Here we see the effect of sacrificing one fingerprint value to indicate empty slots, e.g., for $k=3$, we do not have 8 but only 7 usable fingerprints.
Since for the windowed versions, the fingerprint size is only $k$, this effect is severe.
For the bucketed versions, we have 1 or 2 additional fingerprint bits (for bucket sizes 2 or 4, respectively) and no window offset bits, so the effect is less severe.
While the interplay of these different effects on the FPR for small~$k$ is interesting, it is not so relevant in practice, as most applications aim at FPRs of $2^{-k}$ with $k\ge 5$.

\subsection{Throughput Performance Benchmarks}
\label{sec:eval_time}

\begin{figure*}[t]\centering
\includegraphics[width=1.\textwidth]{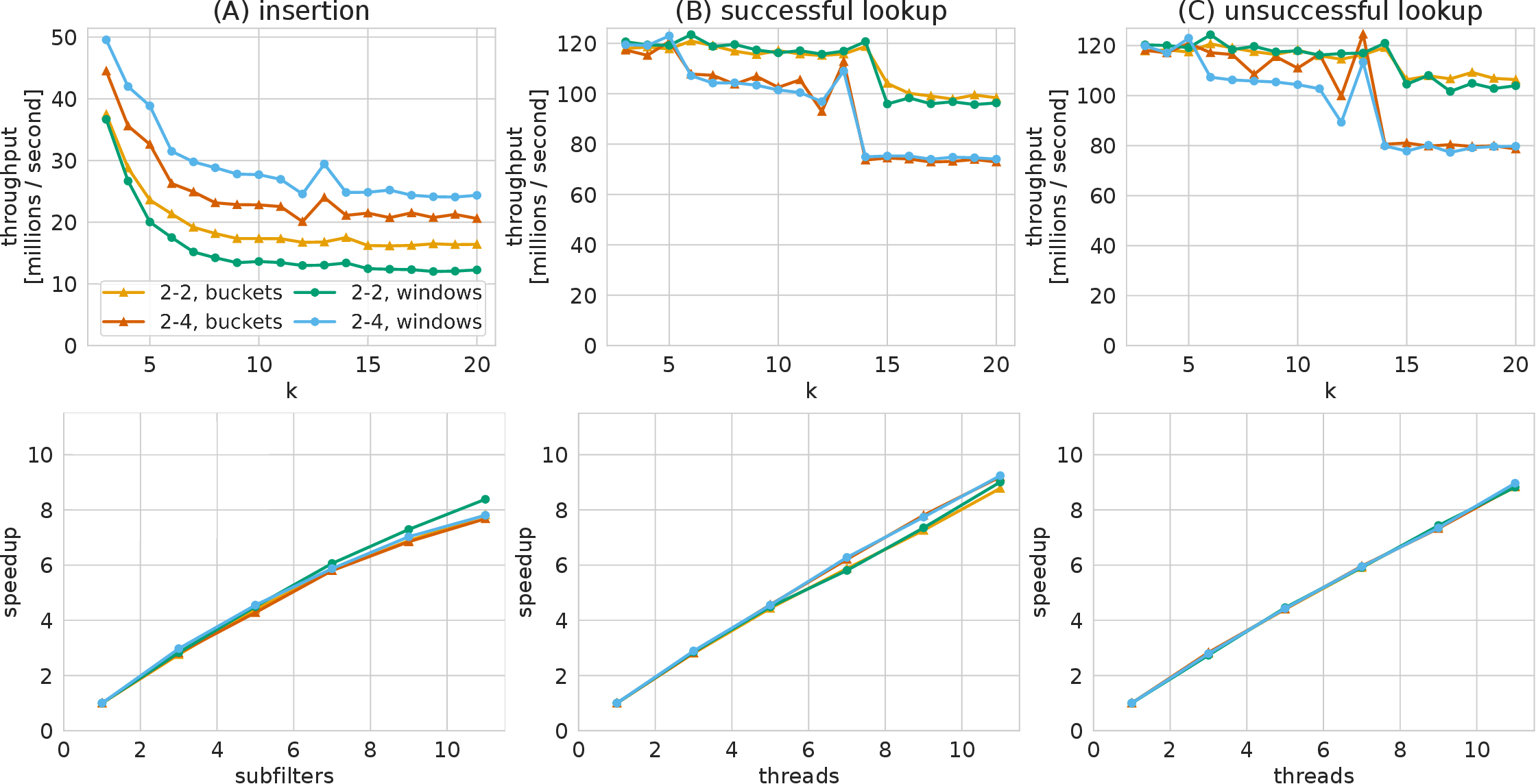}
\caption{
Performance benchmarks for (A) insertions, (B) successful, and (C) unsuccessful lookups of $2 \cdot 10^9$ integer keys, with a maximum random walk length of 10\,000, averaged over five runs.
Top: Throughput in million keys per second for different values of~$k$ using five subfilters (inserter threads) or five query threads; higher throughput is better. 
Bottom: Speedup factor for varying number of threads, for a fixed FPR of $2^{-10}$.
}\label{fig:benchmarks}
\end{figure*}

We evaluate insertion throughput and query throughput separately for successful and unsuccessful queries for different~$k$, with a load factor of 98\% of the load threshold.
Figure~\ref{fig:benchmarks}~(top) shows the throughput in million keys per second (wall time) using five inserter and query threads and different values of~$k$. 
Insertion throughput is lower compared to query throughput. 
This is expected since the random walk may cause several cache misses. 
In contrast, queries incur at most two cache misses.
The $(2,4)$ windowed Cuckoo filters have the highest insertion throughput, and $(2,2)$ windowed Cuckoo filters have the lowest insertion throughput, which is in concordance with the results from Section~\ref{sec:eval_time_memory}.
The throughput of successful and unsuccessful queries is similarly high (Figure~\ref{fig:benchmarks}B and \ref{fig:benchmarks}C).

For $l=4$, the whole bucket or window fits into one 64-bit integer for $k \le 13$. 
In this case, the bit-parallel optimizations described in Section~\ref{sec:engineering} are applied.
For $l=4$ and $k \ge 14$, we check each slot in the bucket or window separately, hence query throughput drops.
For $l=2$, both windows (or buckets) fit into one 64-bit integer for $k \le 14$, and throughput drops for $k \ge 15$.
The increased throughput at $k\in\{5,13\}$ for $l=4$ and at $k\in\{6,14\}$ for $l=2$, is obtained due to the optimized memory access when the number of bits in a slot is a multiple of~8.
Small variations for different values of~$k$ may be caused by different compiler optimizations during just-in-time compilation that can be applied to individual fixed values of~$k$.

The speedup factor from parallelization is almost linear, both for insertions and for queries for up to 5~threads (Figure~\ref{fig:benchmarks} bottom, see Section~\ref{sec:parallelization} for details).
For more threads, the distributing main thread starts to become the bottleneck.
The speedup for insertions is slightly lower than for queries, possibly due to higher memory bandwidth utilization.

\subsection{Software Prefetching Benchmarks}
\label{sec:benchmark_prefecth}

\begin{table}[tb]
\caption{
Query throughput in million keys per second (higher is better) with and without software prefetching, with different prefetching lookahead offsets between 4 and 256 (single threaded).
Times are averaged over five runs. 
All filters have $2^{30}$ slots and the load 98\% of the theoretical load threshold, such that all insertions succeed with a probability close to~1. 
The number of bits per slot is 16 for all configurations, resulting in a false positive rate of $2^{-14}$ for (2,2) bucketed and windowed Cuckoo filters and of $2^{-13}$ for (2,4) bucketed and windowed Cuckoo filters.
}\label{tab:prefetch}
\begin{tabular}{lcccccccc}
\hline
& & \multicolumn{7}{c}{Prefetch offset}\\
Filter & None & 4 & 8 & 16 & 32 & 64 & 128 & 256 \\
\hline
(2, 2) bucketed & 52.00 & 61.62 & 89.26 & 132.77 & 169.26 & 195.92 & \textbf{205.58} & 197.48 \\
(2, 4) bucketed & 44.60 & 58.40 & 85.14 & 120.52 & 133.90 & \textbf{152.23} & 149.11 & 137.57 \\
(2, 2) windowed & 38.49 & 59.32 & 86.15 & 126.32 & 149.04 & \textbf{174.69} & 173.75 & 166.68 \\
(2, 4) windowed & 35.64 & 55.17 & 80.18 & 107.21 & 117.97 & \textbf{129.78} & 125.77 & 117.23 \\
\hline
\end{tabular}
\end{table}

We evaluate the effect of explicit software prefetching on the query throughput on random 64-bit integers (50\% contained and 50\% not contained keys, shuffled).
The results for prefetching the required cache lines 0 to 256 iterations ahead are shown in Table~\ref{tab:prefetch}.
The throughput of all Cuckoo filter variants increases using software prefetching because the number of cache misses is reduced. The optimal prefetch offset on the tested hardware (see Section~\ref{sec:setup}) is 64 for all filters except the $(2,2)$ bucketed Cuckoo filter, where the highest throughput is achieved for an offset of 128. For the optimal offset, the throughput is 3 to 4 times higher for all filters compared to not performing any prefetching.
However, the speedup and the optimal offset highly depends on the hardware; see Table~\ref{tab:prefetch_pike} for the same benchmarks on different hardware, where the optimal offset is 8, and the throughput is only increased twofold.

\subsection{Comparison with Prefix Filters and Vector Quotient Filters}
\label{sec:eval_throughput}

\begin{table*}[t]\centering
\caption{
Insertion and query throughput in million keys per second (higher is better) and memory overhead factor $C$ (lower is better) for different filter types and three different values of $k \in \{8, 13, 14\}$, which are optimal or required for different filter types.
All filters are have $2^{30}$ slots, and the load is selected such that all insertions succeed with a probability of almost 1 (load $0.98\,T$ for Cuckoo filters, $T$ for Prefix filters, $0.92\,T$ for Vector Quotient filters, where $T$ is the theoretical load threshold).
}\label{tab:filter_comparison}
\resizebox{1.0\linewidth}{!}{
\begin{tabular}{l@{~~~~~~}ccc@{~~~~~~}ccc@{~~~~~~}ccc}
\hline
  & \multicolumn{3}{c}{target FPR $2^{-8}$} & \multicolumn{3}{c}{target FPR $2^{-13}$} & \multicolumn{3}{c}{target FPR $2^{-14}$}\\
\hline
  & \multicolumn{2}{c}{throughput} & &  \multicolumn{2}{c}{throughput} & &  \multicolumn{2}{c}{throughput} & \\ 
  filter type & insert & lookup & $C$ &  insert & lookup & $C$ & insert & lookup & $C$ \\
\hline
Cuckoo (2, 2) b & 4.33 & \textbf{44.33} & 1.39 & 4.38 & 39.96 & 1.29 & 4.49 & \textbf{51.98} & 1.28 \\
Cuckoo (2, 4) b & 6.42 & 36.64 & 1.42 & 6.72 & \textbf{44.54} & 1.28 & 5.96 & 22.73 & 1.26 \\
Cuckoo (2, 2) w & 3.10 & 37.10 & \textbf{1.31} & 3.12 & 36.48 & \textbf{1.21} & 3.14 & 38.29 & \textbf{1.20} \\
Cuckoo (2, 4) w & 8.15 & 31.94 & 1.40 & 8.13 & 35.65 & 1.25 & 7.44 & 19.37 & 1.24 \\
Prefix (VQF) & \textbf{33.17} & 39.11 & 1.45 & \textbf{--} & \textbf{--} & \textbf{--} & \textbf{--} & \textbf{--} & \textbf{--} \\
Prefix (CF12) & 32.38 & 39.17 & 1.45 & \textbf{--} & \textbf{--} & \textbf{--} & \textbf{--} & \textbf{--} & \textbf{--} \\
Vector Quotient & 22.10 & 25.44 & 1.48 & \textbf{--} & \textbf{--} & \textbf{--} & \textbf{--} & \textbf{--} & \textbf{--} \\
\hline
\end{tabular}}
\end{table*}

Before comparing the throughput of our Cuckoo filter implementation with other state-of-the-art filters, we note several constraints that complicate a fair comparison.

\begin{enumerate}
\item The Vector Quotient filter (VQF) only works if the number of slots is a power of 2. Hence, we set the number of slots to $2^{30}$. In practice, the space overhead could thus be considerably larger for the VQF. In contrast, our Cuckoo filter implementation and the Prefix filter are flexible concerning the number of slots.
\item Both the VQF and the Prefix filter only support an FPR of $\approx 2^{-8}$ (and $2^{-16}$ for the VQF), and their performance is optimized for this special case. 
A comparable Cuckoo filter with 8~bits per slot has an FPR of $2^{-5}$. 
In contrast, we support all FPRs with different optimizations due to the advantages of just-in-time compilation, but at the cost of being less optimized for one special case.
\item The available Prefix filter implementation only works on CPUs that support AVX512 instructions; there is no fallback implementation.
\item Since the other filter implementations do not provide the option to preload the memory locations for a key, we measure the throughput without software prefetching.
\end{enumerate}

\subparagraph{Throughput}
We evaluated the insertion and lookup throughput (50\% contained keys and 50\% not contained keys, shuffled) on random 64-bit integers (see Table \ref{tab:filter_comparison}).
Prefix filters have the highest insertion throughput. Cuckoo filters have the lowest insertion throughput due to many cache misses during the random walk compared to $\approx 1$ cache miss for Prefix filters and two for Vector Quotient filters.
For $k=8$, the lookup throughput of Prefix, Vector Quotient and Cuckoo filters is similar. In the optimized cases (i.e., Cuckoo filters with 16 bits per slot), bucketed Cuckoo filters have a higher lookup throughput compared to Prefix and VQF filters. 
The lower throughput of windowed vs.\ bucketed Cuckoo filters is explained by the fact that windows are not cache-aligned and a certain fraction of windows spans two cache lines.
For a comparison of query throughput for different types of queries (only contained keys, only not contained keys), see Appendix~\ref{app:query}.

\subparagraph{Space overhead}
The space overhead depends on the FPR and a valid comparison is thus only possible for the same FPR.
For a given FPR, the $(2,2)$~windowed Cuckoo filter has the smallest space overhead compared to Prefix, VQF and bucketed Cuckoo filters (factors~$C$ in Table~\ref{tab:filter_comparison}). 

\subparagraph{Summary}
Our design has a smaller space overhead compared to state-of-the-art filters, while lookup times remain competitive. 
If fast insertion times are essential and an FPR of $2^{-8}$ is tolerable, Prefix filters are preferable.

\section{Conclusion and Discussion}
\label{sec:conclusion}

We introduced improvements of Cuckoo filters, enabling their more flexible use and reducing their overhead factor from $1.05(1+3/k)$ to $1.06(1+2/k)$, by switching from a $(2, 4)$ bucketed layout to a $(2, 2)$ windowed layout.
This is an improvement for relevant FPRs (e.g.\ around $k=10$, up to $k\le 102$).
For an FPR of $2^{-8}$, $(2, 2)$ windowed Cuckoo filters need $\approx 10.6$ GB to index 10 billion keys, compared to $\approx 11.6$ GB for $(2, 4)$ bucketed Cuckoo filters. 
Another interesting aspect of this work is that just-in-time compiled Python code can adjust to run-time specified parameters (such as $k$), resulting in a flexible implementation of Cuckoo filters that is competitive in speed with a \texttt{C++} implementation. For batch queries, software prefetching increases throughput even further.

In comparison to other state-of-the-art filters, our implementation of $(2,2)$~windowed Cuckoo filters has the smallest space overhead and competitively fast lookup times, especially for optimized cases, such as $k=14$, where four 16-bit slots can be examined in a bit-parallel manner in a single 64-bit integer.
In that case, the overhead factor of $1.20$ even outperforms static XOR filters \cite{graf_xor_2020} that need a factor of $1.23$, originally advertised as ``faster and smaller than Bloom and Cuckoo filters''. (Of course, by now, better static filters exist \cite{dillinger_ribbon_2021}.)

For $k=8$, Prefix filters are highly optimized concerning fast insertions, here Cuckoo filters are not competitive in terms of speed. In this case, the increased speed may outweigh the slightly larger space overhead of Prefix filters.
For even lower values of~$k$, Blocked Bloom filters are simpler and can be competitive.

In summary, the low space requirements and fast query times make the improved Cuckoo filters valuable candidates for many applications, e.g., in DNA sequence analysis applications \cite{zentgraf_fast_2021,zentgraf_cleanifier_2025}, where large exact hash tables may be replaced by smaller filters at the cost of a few false positive queries.

\section{Code availability}
The code and workflows to reproduce the results are available at \gitrepo.

\bibliography{sn-bibliography}

\newpage\appendix
\part*{\sffamily Appendix}

\section{Distribution of random walk lengths}
\label{app:random_walk_distribution}

\begin{figure}[t]\centering
\includegraphics[width=1.\linewidth]{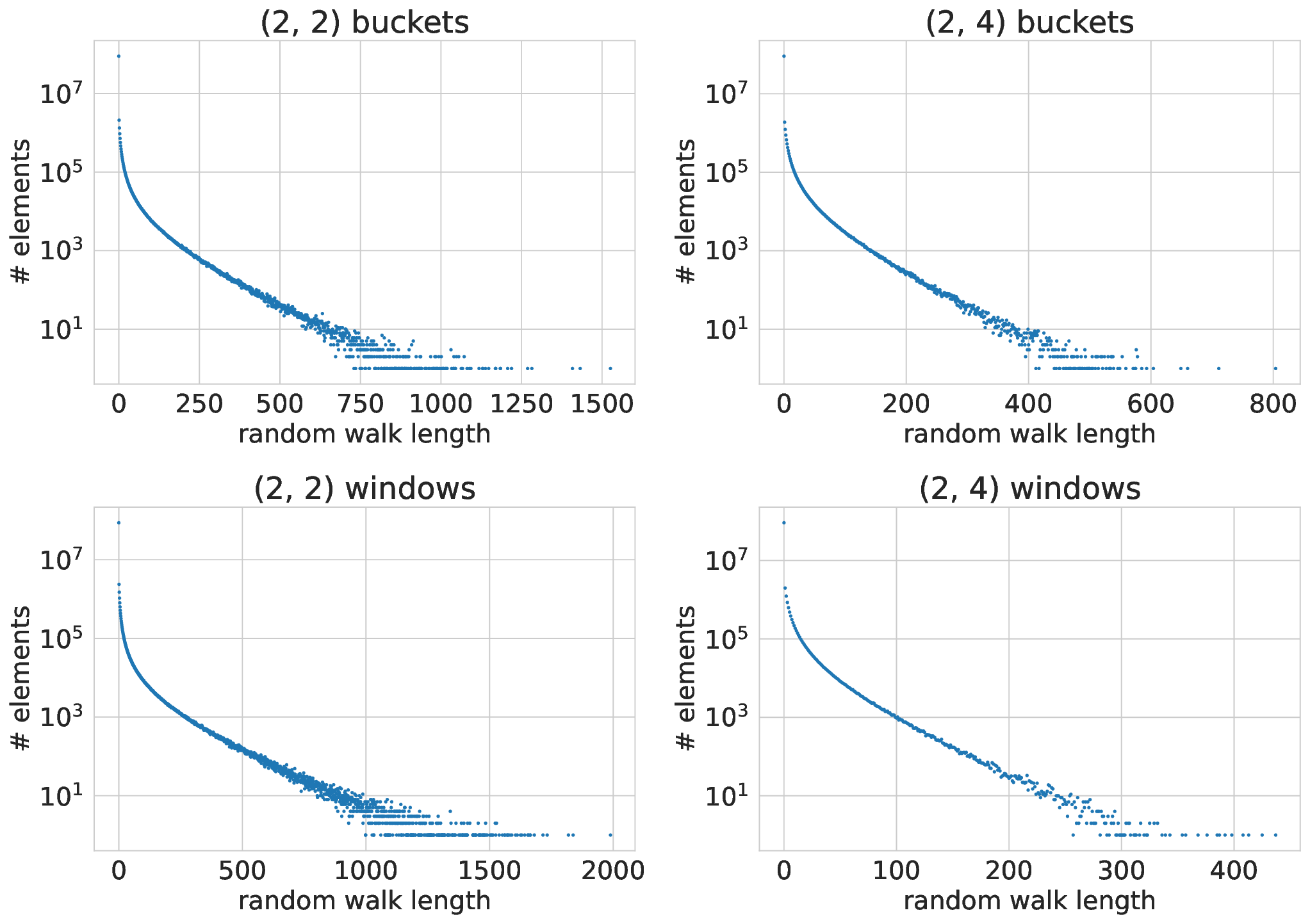}
\caption{
Distribution of random walk lengths when inserting $10^8$ elements into different Cuckoo filter types for a fixed FPR of $2^{-10}$ and a filter size that ensures a load of at most $98\%$ of the theoretical load threshold.
}\label{fig:random_walk_histogram}
\end{figure}

Long random walks cause many cache misses and hence lead to slow insertion times.
Therefore, short random walks are preferable.
This can easily be achieved by increasing the space overhead of the filter, since the costly insertions happen when the filter is almost filled to its maximum capacity (see Figure \ref{fig:max_fill}).
However, in many applications small filters are preferable, in particular, if more queries than insertions are performed.

Figure \ref{fig:random_walk_histogram} shows the random walk length distribution of bucketed and windowed Cuckoo filters allowing a maximum random walk length of $10^4$. 
Most keys are directly inserted into a free slot.
Only for a small fraction of keys does the filter perform a long random walk; note the logarithmic scale of the y-axis. 
Choosing a maximum load such that the filter is only filled up to $98\%$ of its maximum capacity ensures that the actual maximum random walk length is much lower than the allowed maximum random walk length of $10^4$, with $(2, 4)$ Cuckoo filters having shorter random walks compared to $(2, 2)$ Cuckoo filters.

\section{C++ vs.\ Python/numba implementation}
\label{app:numba}

To rule out that our Cuckoo filter implementation is slower or faster due to the use of just-in-time compiled \texttt{Python}, we compared the runtime of the original implementation of the Cuckoo filter (a~$(2,4)$-bucketed Cuckoo filter, computing the alternative bucket with an \texttt{XOR} operation, implemented in \texttt{C++} \cite{fan_cuckoo_2014}) with our implementation of the the same (2,4)-bucketed Cuckoo filter. 
Both implementations are as similar as possible, e.g., both implementations use the same trick for bit-parallel lookups, but for example, different hash functions.

Table \ref{tab:cmp_c++_numba} shows the throughput for both implementations of Cuckoo filters at 95\% load, averaged over three repeated runs. 
We measure lookup time by querying at 50:50 mixture of keys that are contained and not contained in the filter.
Since both implementations have comparable throughput, in particular for lookups, the \texttt{numba} implementation does not affect our reasoning on an algorithm engineering level when comparing our novel Cuckoo filters with other state-of-the-art probabilistic filters.

\begin{table}[t]\centering
\caption{
Throughput in million keys per second (higher is better), single threaded.
We compare the original Cuckoo filter implementation written in \texttt{C++} with an implementation in just-in-time compiled \texttt{Python} using \texttt{numba}.
All filters have $2^{30}$ slots and a load of $95\%$.
}\label{tab:cmp_c++_numba}
\begin{tabular}{l@{~~~~~~}cc@{~~~~~~~~~~~~}cc}
\hline
  & \multicolumn{2}{@{}l}{FPR $2^{-5}$ (8-bit slots)} &  \multicolumn{2}{@{}l}{FPR $2^{-13}$ (16-bit slots)} \\
  & insert & lookup & insert & lookup \\ \hline
\texttt{C++} & 7.305 & 61.939 & 7.371 & \textbf{60.517}\\
\texttt{numba} & \textbf{8.938} & \textbf{70.688} & \textbf{9.104} & 55.013\\
\hline\\
\end{tabular}
\end{table}

\section{Software prefetching}
\label{app:prefetch}

Table~\ref{tab:prefetch_pike} shows the throughput in million keys per second using different prefetch lookahead offsets.
The benchmarks were run on an Ubuntu (24.04 LTS) server with two AMD EPYC 9534 64-Core Processors, 1.5TB of 4800-MHz DDR5 memory and a KIOXIA CD8P SSD).
In contrast to Table~\ref{tab:prefetch}, the throughput is lower for all prefetch lookahead offsets due to the slower RAM and lower CPU clock speed. 
In addition, the throughput increases less, and the best offset is lower (8 in comparison to 128).

\begin{table}[t]\centering
\caption{
Throughput in million keys per second (higher is better) with and without software prefetching using prefetching offsets between 4 and 256, single threaded.
All filters have $2^{30}$ slots with load at 98\% of the theoretical load threshold.
The number of bits per slot is 16 for all configurations, resulting in a false positive rate of $2^{-14}$ for  (2, 2) bucketed/windowed Cuckoo filters,  and of $2^{-13}$ for (2, 4) bucketed/windowed Cuckoo filters.}
\begin{tabular}{lcccccccc}
\hline
& & \multicolumn{7}{c}{prefetch lookahead offset}\\
filter type & none & 4 & 8 & 16 & 32 & 64 & 128 & 256 \\
\hline
Cuckoo (2, 2) bucketed & 36.21 & 42.49 & \textbf{51.50} & 45.05 & 44.82 & 45.16 & 45.87 & 45.31 \\
Cuckoo (2, 4) bucketed & 31.73 & 41.45 & \textbf{54.08} & 45.70 & 45.97 & 46.27 & 47.25 & 46.55 \\
Cuckoo (2, 2) windowed & 26.20 & 38.48 & \textbf{52.71} & 42.83 & 42.56 & 42.96 & 43.08 & 42.02 \\
Cuckoo (2, 4) windowed & 22.87 & 36.79 & \textbf{47.38} & 44.96 & 45.73 & 44.68 & 44.62 & 43.24 \\
\hline
\end{tabular}
\label{tab:prefetch_pike}
\end{table}

\section{Query throughput comparison}
\label{app:query}

Table~\ref{app:filter_comparison} shows the query throughput of the different Cuckoo filter variants, the Prefix filter, and the Vector Quotient filter for different types of queries: querying only contained keys, only not contained keys or a randomly shuffled mix of contained and not contained reads (50/50).
Prefix filters and Vector Quotient filters have a higher throughput when querying not contained keys. However, the throughput of mixed queries is similar to the throughput for contained keys, possibly due to a worse branch prediction. 
The query throughput of the Cuckoo filter variants is similar for all types of queries.

\begin{table*}[t]\centering
\caption{
Query throughput in million keys per second (higher is better) for different filter types and three different values of $k \in \{8, 13, 14\}$, which are optimal or required for different filter types.
Throughput is measured for successful lookups (contained keys; cont.), unsuccessful lookups (not contained keys; not cont.) and 50:50 mixtures (randomly shuffled 50\% contained and 50\% not contained keys).
All filters are have $2^{30}$ slots, and the load is selected such that all insertions succeed with a probability of almost 1 (load $0.98\,T$ for Cuckoo filters, $T$ for Prefix filters, $0.92\,T$ for Vector Quotient filters, where $T$ is the theoretical load threshold).
}\label{app:filter_comparison}
\resizebox{1.0\linewidth}{!}{
\begin{tabular}{l@{~~~~~~}ccc@{~~~~~~}ccc@{~~~~~~}ccc}
\hline
  & \multicolumn{3}{@{}l}{target FPR $2^{-8}$ throughput} & \multicolumn{3}{@{}l}{target FPR $2^{-13}$ throughput} & \multicolumn{3}{@{}l}{target FPR $2^{-14}$ throughput}\\
filter type & cont. & not cont. & shuffled &  cont. & not cont. & shuffled & cont. & not cont. & shuffled \\
\hline
Cuckoo (2, 2) b & \textbf{44.15} & 43.89 & \textbf{44.32} & 39.69 & 39.89 & 39.96 & \textbf{51.59} & \textbf{51.96} & \textbf{51.98} \\
Cuckoo (2, 4) b & 36.72 & 38.41 & 36.64 & \textbf{43.45} & \textbf{47.79} & \textbf{44.54} & 21.20 & 24.25 & 22.73 \\
Cuckoo (2, 2) w &  37.01 & 37.05 & 37.10 & 36.31 & 36.46 & 36.48 & 38.26 & 38.37 & 38.29 \\
Cuckoo (2, 4) w &  31.65 & 33.77 & 31.94 & 35.35 & 38.24 & 35.65 & 19.26 & 20.34 & 19.37 \\
Prefix (VQF) & 36.75 & 53.55 & 39.11 & \textbf{--} & \textbf{--} & \textbf{--} & \textbf{--} & \textbf{--} & \textbf{--} \\
Prefix (CF12) & 38.29 & \textbf{57.37} & 39.17 & \textbf{--} & \textbf{--} & \textbf{--} & \textbf{--} & \textbf{--} & \textbf{--} \\
Vector Quotient &  25.24 & 32.62 & 25.44 & \textbf{--} & \textbf{--} & \textbf{--} & \textbf{--} & \textbf{--} & \textbf{--} \\
\hline
\end{tabular}}
\end{table*}

\end{document}